\documentclass[aps,pre,twocolumn,superscriptaddress]{revtex4}
\usepackage{amsmath,braket}
\usepackage{amsfonts}
\usepackage{graphicx}
\usepackage{hyperref}
\usepackage{verbatim}
\usepackage{mathrsfs} 
\usepackage{cleveref}
\usepackage{xcolor}

\usepackage{CJK}
\hypersetup{
  colorlinks   = true, %Colours links instead of ugly boxes
  urlcolor     = blue, %Colour for external hyperlinks
  linkcolor    = blue, %Colour of internal links
  citecolor   = blue %Colour of citations
}
\usepackage{epstopdf}
\def\be{\begin{equation}}
\def\ee{\end{equation}}
\def\ba{\begin{eqnarray}}
\def\ea{\end{eqnarray}}
\def\he{\hbar_{\rm eff}}
\def\ll{l}
\def\tt{\theta}

\newcommand{\avera}[1]{\left\langle #1\right\rangle}
\newcommand{\norm}[1]{\lVert#1\rVert}
\newcommand{\px}{\mathcal{X}}
\newcommand{\pp}{\mathcal{P}}

\newcommand{\var}{\mathrm{Var}}

\DeclareMathOperator{\Tr}{Tr}

\begin{document}
\begin{CJK}{UTF8}{gbsn}
\title{Quantum ergodicity and mixing and their classical limits with quantum kicked rotor}

\author{Jialong Jiang (姜家隆)}
\affiliation{Yuanpei College, Peking University, Beijing 100871, China}
\author{Yu Chen (陈宇)}
\affiliation{Center for Theoretical Physics, Department of Physics, Capital Normal University, Beijing 100048, China}
\author{Biao Wu (吴飙)} \email{wubiao@pku.edu.cn}
\affiliation{International Center for Quantum Materials, School of Physics, Peking University, Beijing 100871, China}
      \affiliation{Collaborative Innovation Center of Quantum Matter, Beijing 100871, China}
      \affiliation{Wilczek Quantum Center, School of Physics and Astronomy, Shanghai Jiao Tong University, Shanghai 200240, China}
\date{\today}

\begin{abstract}
We study the ergodicity and mixing of quantum kicked rotor (QKR) with two distinct approaches. 
In one approach, we use the definitions of quantum ergodicity and mixing recently proposed in [Phys. Rev. E 94, 022150 (2016)], 
which involve only eigen-energies (Floquet quasi-energies for QKR).  
In the other approach, we study ergodicity and mixing with quantum Poincar\`e section, 
which is plotted with a method that maps a wave function unitarily onto quantum phase space composed of Planck cells. 
Classical Poincar\`e section can be recovered with the effective Planck constant gradually diminishing. 
We demonstrate that the two approaches can capture the quantum and classical characteristics of ergodicity and 
mixing of QKR, and give consistent results with classical model at semiclassical limit.
Therefore, we establish a correspondence between quantum ergodicity (mixing) and classical ergodicity (mixing). 
\end{abstract}
\maketitle
\end{CJK}

\section{Introduction}

Ergodicity and mixing in classical dynamics are essential to the foundation 
of classical statistical mechanics~\cite{EH}. 
Ergodicity enables an isolated classical system to equilibrate dynamically 
whereas mixing ensures that the fluctuation is small at equilibrium. 
As the dynamics of microscopic particles is described by quantum mechanics, 
one would naturally want to generalize ergodicity and mixing to quantum dynamics 
to set up the foundation of quantum statistical mechanics.
However, such a generalization faces two apparent obstacles: (1) quantum 
dynamics is inherently linear while ergodicity and mixing are
intrinsically related to the nonlinear and chaotic nature of classical dynamics~\cite{EH}; 
(2) quantum dynamics is described by states in Hilbert space rather than trajectories.  

Nevertheless, many have attempted to introduce ergodicity and/or mixing to quantum dynamics. 
To our knowledge, von Neumann was the first to discuss ergodicity in quantum dynamics in a 1929 paper~\cite{Neumann1929,von2010proof}, where he proved quantum ergodic theorem. 
For some not well understood reasons, 
this work of von Neumann did not receive enough attention for a long time~\cite{goldstein}. 
In 1984, Peres made another attempt to define ergodicity and mixing for quantum 
dynamics~\cite{peres1984,peres2}, where he had to first define the concept of quantum chaos in a specific way
that has not been widely accepted. It is not clear that how Peres' definitions can be generalized to spin systems. 
 
Quantum ergodicity and mixing were recently defined in a different way in Ref.~\cite{zhang2016ergodicity}. 
These definitions are inspired by von Neumann's 1929 work, and use only eigen-energies of a given quantum system: 
({\it i}) the quantum system is ergodic if  its eigen-energies are not degenerate.
({\it ii}) the quantum system is mixing if there is no degeneracy in the differences between any pair of its eigen-energies. 
 Such definitions, which can be readily applied to spin systems,  
 are mathematically rigorous and lead to quantum dynamics with the properties we need: 
(1) with condition ({\it i}), the observables for a typical quantum state equals to the corresponding ensemble average, which is a manifestation of ergodicity; (2) with condition ({\it ii}), the fluctuations of observables around  their long-time average are relatively small, which corresponds to mixing.
For convenience, we shall refer to such definitions of ergodicity and mixing as eigen-energy (EE) definitions.  

In this work, we study quantum ergodicity and mixing with quantum kicked rotor~\cite{CasatiBook, ChirikovScholarpedia, FishmanScholarpedia, Fishman82}.  We use two different approaches, one involving EE definitions of quantum ergodicity and mixing and the other employing quantum Poincar\`e section. They lead to
two sets of  results, which  are consistent with each other. In the semiclassical limit, $\hbar\rightarrow 0$,  these quantum results
are also consistent with classical ergodicity and mixing. 

In the first approach we compute numerically the Floquet quasi-energies of QKR, and two parameters 
$\eta$ and $\zeta$ that characterize the degeneracy of these Floquet quasi-energies and 
their pair-wise difference, respectively~\cite{zhang2016ergodicity}.  
We find that both parameters have a sharp drop around kicking strength $ K = 1 $, which agrees very well 
with the critical kicking strength $ K_c = 0.972 $ where the classical kicked rotor  starts to become 
chaotic~\cite{CasatiBook, ChirikovScholarpedia}.

In the second approach we project unitarily a wave function to quantum phase space, which is obtained 
by dividing classical phase space into Planck cells. This allows us to observe the dynamical 
evolution of  QKR in quantum phase space, in a fashion very similar to classical Poincar\'e section. 
The quantum Poincar\'e section begins to appear ergodic and mixing when the kicking strength becomes larger than $K_c$.
Furthermore, we can  prove that the kicked rotor initially localized in a Planck cell
will evolve like its classical counterpart when the effective Planck constant  goes to zero. 
This is illustrated numerically by the striking similarity between the quantum Poincar\'e section
and classical Poincar\'e section of kicked rotor. 
Our results imply that the rather abstract EE definitions of quantum ergodicity and mixing 
are intimately related to the usual intuitive understanding of classical ergodicity and mixing. 

\section{Kicked Rotor}

A quantum kicked rotor (QKR) describes a quantum particle moving in a ring with periodical kicking~\cite{CasatiBook, ChirikovScholarpedia, FishmanScholarpedia, Fishman82, Tian10}.
The Hamiltonian of the  quantum kicked rotor is
\be
\mathcal{H}(\ll,\tt,t)= \frac{\ll^2}{2I} + K\frac{I}{\tau} \cos \tt\sum_{n=-\infty}^\infty \delta(t-n\tau)\,, 
\label{ham}
\ee
where we have $I$ for the moment of inertia, $\tau$ for the kicking period, and $K$ for the kicking strength.  
With $\tau$ as the unit of time, the QKR obeys the following dimensionless Schr\"odinger equation
\be
i\he \frac{\partial}{\partial t}\psi=-\frac{\he^2}{2}\frac{\partial^2}{\partial \tt^2}\psi+
K[\cos \tt\sum_{n=-\infty}^\infty \delta(t-n)]\psi\,,
\ee
where $\he=\hbar\tau/I$ is the effective Planck constant. The Hamiltonian (\ref{ham})
describes a classical kicked rotor (CKR) when $\tt$ and $ \ll $ are treated as classical quantities.
With large kicking strength, the corresponding CKR becomes chaotic or specifically, ergodic and mixing. 

We use QKR to illustrate  quantum ergodicity 
and mixing. As QKR along with its classical counterpart is very well studied, we are able to take advantage of many interesting results in literature~\cite{CasatiBook, ChirikovScholarpedia, FishmanScholarpedia, Fishman82, Tian10}. 
In particular, QKR was already used to show the connection between quantum mixing and its classical counterpart in Ref.~\cite{Toda1988PRL,Toda1989}. 
However, the authors in Ref.~\cite{Toda1988PRL,Toda1989} did not offer a clear definition of quantum mixing and put their emphasis on the effects of external noise.  In this work, we examine the EE definitions of 
quantum ergodicity and mixing with QKR, and show that 
they are consistent with  the intuitive understanding of classical ergodicity and mixing in the limit $\he \to 0$. 

In general, there are two important time scales in QKR 
with a fixed $\he$. The first time scale is Ehrenfest time $t_\hbar$, below which the system is  in the classical region~\cite{Larkin68,Berman78a,Berman78b,Zaslavsky81review,Izrailev90}. 
In the chaotic region,  $t_\hbar\sim\left|\ln(A/\he)\right|/\lambda_L$~\cite{Berman78a,Berman78b,Zaslavsky81review,Izrailev90},
where $A$ is the area in phase space and $\lambda_L$ is the Lyapunov exponent. 
In this case, the Ehrenfest time $t_\hbar$ is very small. Another important time scale is the Heisenberg time $t_H$, 
beyond which the system is pure quantum in localization phase~\cite{FishmanScholarpedia, Fishman82} or super-metal phase~\cite{Izrailev90,Fishman03,Wimberger11,Tian10} depending on whether $\he/4\pi$ is irrational. 
As $t_H \propto 1/\he^2$~\cite{Izrailev90}, the Heisenberg time $t_H$ is usually large.  In our study,
we focus on the time scale $t_\hbar<t<t_H$, where the system is in a quantum-classical crossover region.
Near the end, we will briefly discuss the case $t\gtrsim t_H$, and show  that quantum corrections can affect 
the quantities that we define to measure quantum ergodicity and mixing.

In our study we choose that $\he=2\pi/N$ with $N$ being a large positive integer.
$ N $ measures the system size in $ k $ space, and also defines the time scale of our study. 

\subsection{Construction of quantum phase space}

\begin{figure*}[ht]
\includegraphics[width=0.75 \linewidth]{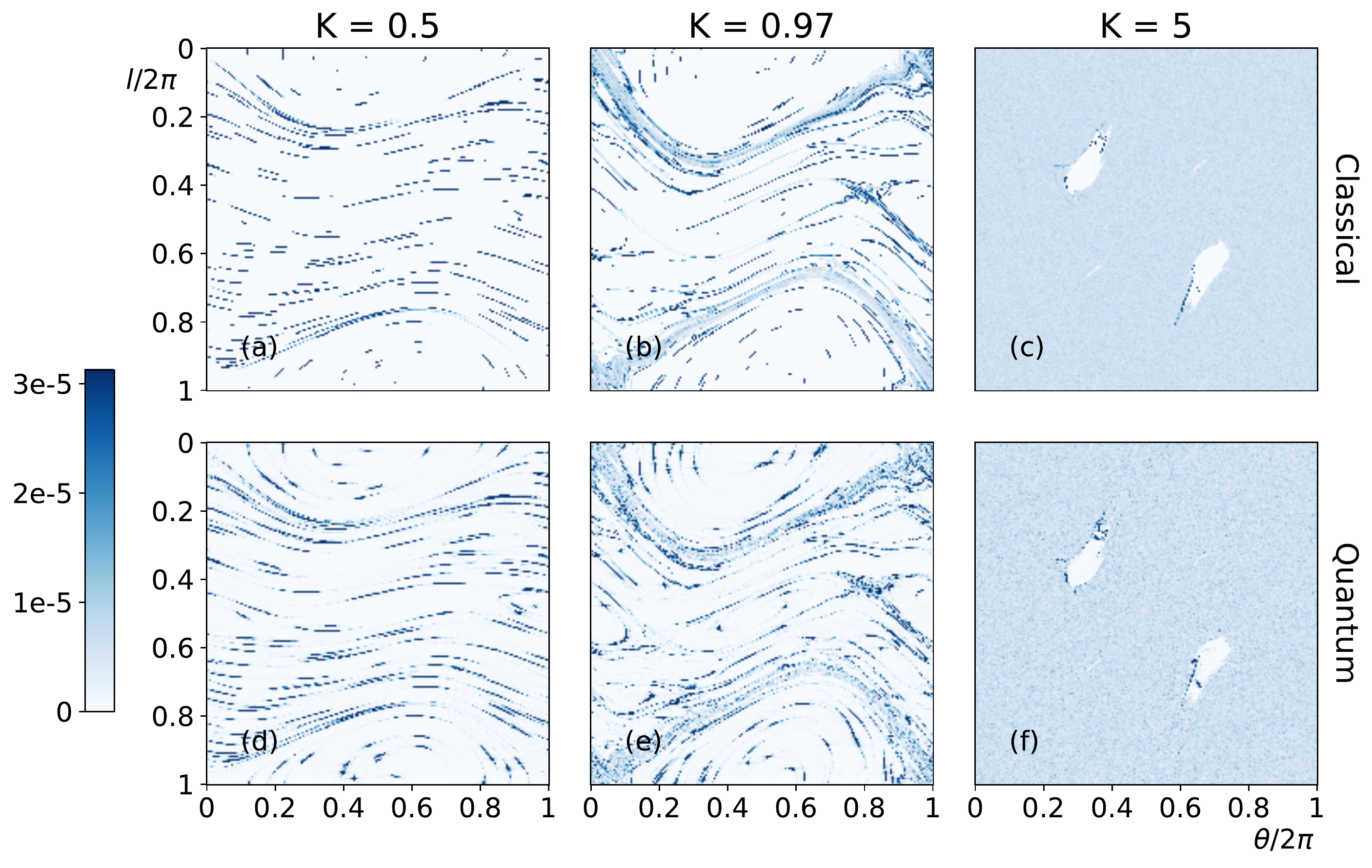}
\caption{(color online) Comparison of the quantum and classical Poincar\'e sections of kicked rotor. 
To make comparison, the classical phase space is  coarse-grained to the same gird of Planck cells. 
The initial state of quantum Poincar\'e section is the superposition of randomly chosen $\ket{\px,\pp}$ states with the same phase. 
We sum the probability of Planck cells whose $ \pp $ differ by the multiple of $ N $, which 
correspond to the period representation on $ l $ in CKR.
The classical Poincar\'e section is an ensemble of particles that have the same distribution 
with the quantum initial states. $\he=2\pi/N$, $N=\ell^2$, and $\ell=400$. }
\label{phase}
\end{figure*}

To facilitate our study, we construct  the quantum phase space by dividing classical phase space 
into Planck cells then project a quantum state onto it unitarily. 
This method was first proposed by von Neumann~\cite{von2010proof} and has recently been developed
in Ref.~\cite{han2015entropy,Fang2017}. The primary advantage of this method
is that as the projection is unitary it gives us a true probability distribution for a quantum 
state in phase space. As a result, we are able to define quantum entropy over phase space~\cite{han2015entropy} and plot quantum Poincar\`e section (see Fig.~\ref{phase}).  
The traditional methods such as Wigner function~\cite{wigner1932quantum}, P representation~\cite{Sudarshan,Glauber}, and Q representation~\cite{Husimi}, along with 
the recent biorthogonal method~\cite{Tannor2012PRL,Larsson2016,Tannor2016}, can only give us
quasi-probabilities. 

In Ref.~\cite{han2015entropy,Fang2017},  the basis wave functions used for unitary projection
are obtained numerically by orthonormalizing a set of Gaussian wave functions. 
For QKR we choose a different set of basis wave function, which is constructed 
analytically with a superposition of finite 
momentum eigenstates (SFME). Many properties, such as localization, of these SFME basis wave functions
are discussed in Appendix~\ref{app_wave_packet}. Here we only show the construction procedure. 

As $\tt$ is of period $2\pi$, the momentum eigenstate has the form 
$ \ket{n} = e^{in\tt} / \sqrt{2\pi} $ with wave number $n$ being an integer.  
We define angle and momentum translation operators as
\ba
\hat{T}_\tt(\px \Delta\tt)\ket{n} &=& \exp(-i n \px \Delta\tt)\ket{n}, \\
\hat{T}_l(\pp \Delta \ll)\ket{n} &=& \ket{n+ \pp\Delta \ll/\he},
\ea
where $\px$ and $\pp$ are integers, $\Delta\tt$ and $\Delta \ll$ are displacements in $\theta$ space and $\ll$ space respectively. 
With a given positive integer number $\ell$, we  start with the wave function
\be
%\ket{\phi_\ell} 
\ket{0, 0} = \frac{1}{\sqrt{\ell}}\sum_{n = 1}^{\ell}\ket{n}\,, 
\ee
which is localized in both angle $\tt$ and angular momentum $\ll$ (see Appendix~\ref{app_wave_packet}).  
By these translation operators, we can then construct a set of  basis as follows
\ba
\ket{\px,\pp} \equiv \hat{T}_\theta(\px \Delta \theta)\hat{T}_l(\pp \Delta l)|0,0\rangle \,,
\label{xp}
\ea
where $\Delta \ll = \ell\he$ and $\Delta \tt= 2\pi/\ell$. Notice that $\theta\in[0,2\pi)$, therefore $\px =0,1,\cdots,\ell-1$. These bases are orthonormal and complete, that is, $\langle \px'.\pp'|\px,\pp\rangle=\delta_{\px'\px}\delta_{\pp'\pp}$ and $\sum_{\px=0}^{\ell-1}\sum_{\pp}|\px,\pp\rangle\langle\px,\pp| = 1$.  

With this construction we obtain a quantum phase space, which consists of a series of Planck cells numbered by 
two integers $\px$ and $\pp$. Each Planck cell is assigned a localized wave function $|\px,\pp\rangle$.
One can project any wave function $\ket{\psi}$ to this phase space unitarily as 
$\ket{\psi}=\sum  \ket{\px,\pp} \braket{\px,\pp|\psi}$, and $P_{\px,\pp}=|\braket{\px,\pp|\psi}|^2$ is 
the probability at Planck cell $(\px, \pp)$.   
More details about this set of basis can be found in Appendix~\ref{app_wave_packet}. Note
that this SFME basis was used in Ref.~\cite{Toda1988PRL,Toda1989} to examine the noise effects
on the dynamics of QKR. 

In our QKR study, for the effective Planck constant $\he=2\pi/N$,  we choose $N=\ell^2$. 
In this way, we get a balanced resolution for $\tt$ and $\ll$ as  the number 
of Planck cells along the $\tt$ direction and the one along the $\ll$ direction are both $\ell$.

%%%%%%%%%%%%%%%%%%%%%
%%%%%%%%%%%%%%%%%%%%%

\subsection{Quantum-classical correspondence of kicked rotor}

For a CKR, its equations of motion can be represented as
the Chirikov standard map that connects the momenta and positions after the $(j-1)$th and the $j$th kicks~\cite{CasatiBook, ChirikovScholarpedia}, 
\ba
\label{ckr1}
\tt_j  &=&  \tt_{j-1}+\ll_{j-1}\,, \\ 
\ll_j  &=& \ll_{j-1}+K\sin \tt_{j}\,,
\label{ckr2}
\ea
where the angular momentum $\ll_j$ is scaled with $I/\tau$. As the particle lives on a ring, $\tt_j$ is clearly periodic. The momentum $\ll_j$ can also be regarded as periodic since $\ll_j+2n\pi$ means that the particle rotates $n$ more rounds while not affecting how the momentum changes in the next kick. 
Therefore, the Poincar\'e section of CKR is always presented with periodical boundary condition both in angle and angular momentum~\cite{CasatiBook, ChirikovScholarpedia} (see also Fig. \ref{phase}). 

For QKR,  we consider the map connecting the momentum eigenstates immediately after the adjacent kicks. The transition matrix element from $ \ket{n} $ to $ \ket{m}$ is
\be
\hat{U}_{m,n} = \braket{m|\hat{U}|n}=J_{m-n}(K/\he)e^{-in^2\he/2}\,,
\ee
where $ J_n(K/\he) \equiv \mathbb{J}_n(K/\he)/i^n $ with $ \mathbb{J}_n(K/\he) $ being the first kind Bessel function. 
With this transition matrix, one can compute how the wave function of a QKR changes after each kick. 
Using the SFME basis in the last subsection, we can project the wave functions to the quantum phase space
and plot the quantum Poincar\'e section for QKR. The results are shown in Fig.~\ref{phase} and compared
to the corresponding classical Poincar\'e sections. We observe striking similarity between
them, indicating that the quantum dynamics given by $\hat{U}$ can be reduced to the standard map in Eq.~(\ref{ckr1}, \ref{ckr2}) at the limit $\he\rightarrow 0$. In the following, we show analytically that this is indeed the case.

We consider how QKR evolves dynamically when it starts at $\ket{\px_0,\pp_0}$. This corresponds to the CKR starting around ${\tt_0\approx 2\pi\px_0/\ell, \ll_0\approx \pp_0 \ell\he}$.  After one kick, the state becomes $\hat{U}\ket{\px_0,\pp_0}$. Using 
\be
\mathbb{J}_n(K/\he)/i^n = \frac{1}{2\pi} \int_0^{2\pi} e^{i(nm-K\cos m /\he)} dm
\ee
we find that the probability of the state at $ \ket{\px,\pp} $ is
\begin{widetext}
\be
\braket{\px, \pp|\hat{U}| \px_0, \pp_0}= \frac{1}{2\pi\ell}\sum_{\alpha = \ell\pp+1}^{\ell(\pp+1)}
\sum_{\beta= \ell\pp_0+1}^{\ell(\pp_0+1)} 
\int_0^{2\pi}dm
\exp \left\{ \frac{i\ell^2}{2\pi} (
\frac{2\pi\px}{\ell}\frac{\alpha}{N}
-\frac{2\pi\px_0}{\ell}\frac{\beta}{N}
-\pi\frac{\beta^2}{N^2}
+(\frac{\alpha}{N}-\frac{\beta}{N}) m
-\frac{K}{2\pi}\cos m
)\right\}\,.
\ee
\end{widetext}
The above summation can be approximated by integration
\be
P_{\px,\pp} \approx \frac{\ell^3}{2\pi} 
\int_{\ll}^{\ll+\frac{2\pi}\ell}\mkern-15mu d\bar{\alpha}
\int_{\ll_0}^{\ll_0+\frac{2\pi}\ell}\mkern-15mu d\bar{\beta}
\int_{0}^{2\pi}\mkern-15mu dm\,
\exp{\frac{i\ell^2}{2\pi}f(\bar{\alpha},\bar{\beta}, m)}\,.
\ee
where 
\be
f(\bar{\alpha}.\bar{\beta},m)=\bar{\alpha}\tt-\frac{\bar{\beta}^2}{2}-
\bar{\beta}\tt_0+(\bar{\alpha}-\bar{\beta})m -K\cos m\,.
\ee
In the above, we used 
$\tt=\px\Delta \tt$, $\ll=\pp\Delta \ll$, $\bar{\alpha}=\alpha\he$, $\bar{\beta}=\beta\he$. 
According to the method of steepest decent, in the limit $ \ell \rightarrow \infty$, the above probability is non-zero if and only if the partial derivatives of the function $ f(\bar{\alpha},\bar{\beta},m) $ vanish. 
This leads us to the standard map for the CKR.
\ba
\tt &=& \tt_0+\ll_0\,,\\
\ll &=& \ll_0+K\sin \tt\,.
\ea
The above analytical result shows that the dynamics of QKR can be reduced to the classical dynamics 
in the limit of  $\he\rightarrow 0$ or $N=\ell^2\rightarrow \infty$. 
This can be interpreted as the dynamics goes to classical as the scale of the system (moment of inertia $ I $ in our case) 
becomes macroscopic. This is indeed what 
have observed numerically in Fig.~\ref{phase}.  Such a correspondence implies that QKR 
should possess similar dynamical properties of CKR, for example,  if
CKR is ergodic and mixing, QKR intuitively should also be ergodic and mixing. 
This is the focus of the next section.  

\section{Ergodicity and mixing in quantum kicked rotor}
%%%%%%%%%%%%%%%%%%%%%
\begin{figure*}
\includegraphics[width=0.85 \linewidth]{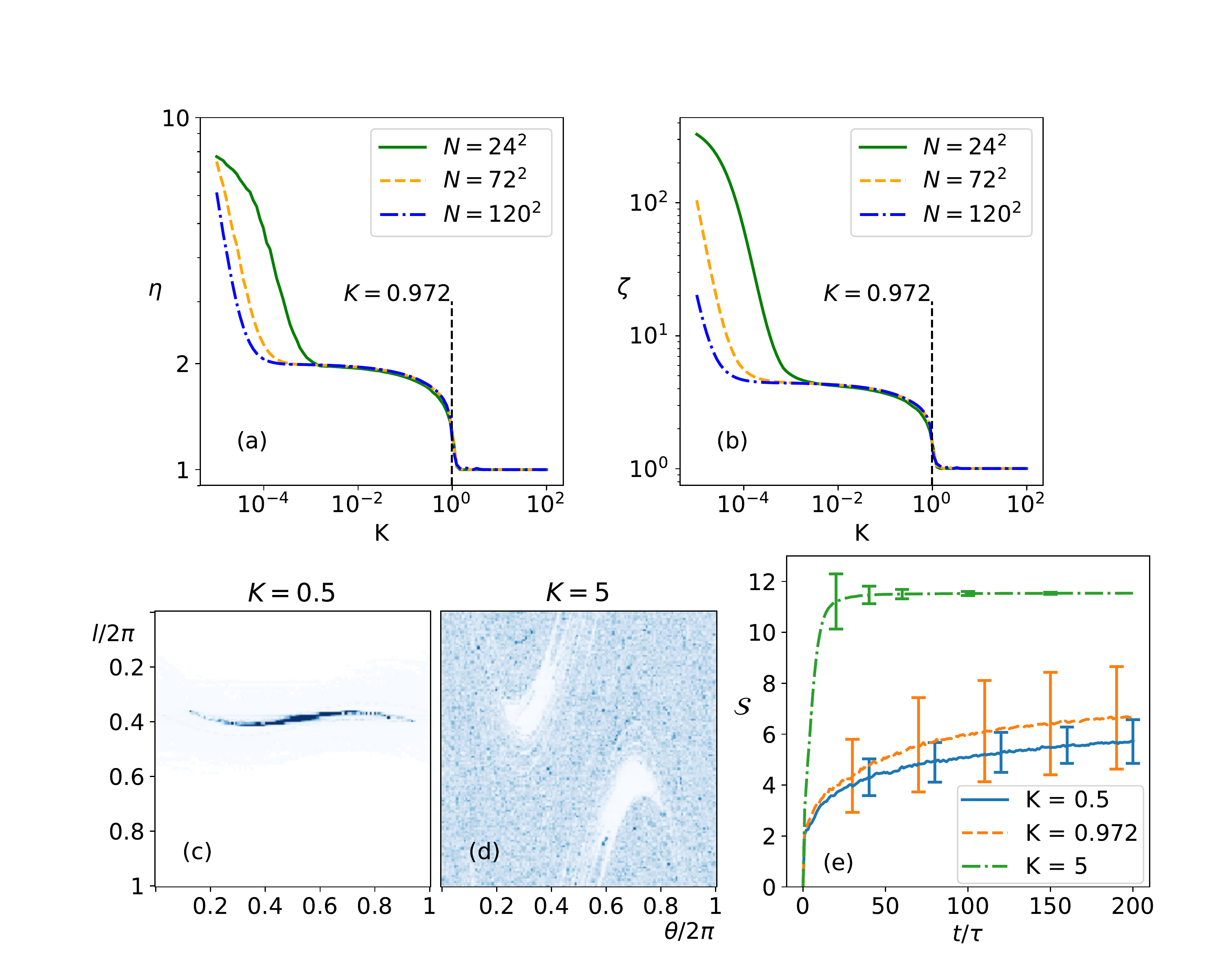}
\caption{(color online) (a)(b) The change of $ \eta $ and $\zeta$ with $K$ at different $ N $. 
(c)(d) The quantum state after 14 kicks with the initial state being a single Planck cell. $ \ell = 140 $
(e) The time evolution of entropy $ \mathcal{S} $ for $ \ell = 400 $ starting from a single Planck cell.
Lines are the average over different starting Planck cells, and the error-bars are the standard deviations. $\he=2\pi/\ell^2$, $N=\ell^2$.  }
\label{comp}
\end{figure*}
%%%%%%%%%%%%%%%%%%%%%
In classical mechanics, ergodicity means the system evolves to almost every points in phase space given enough time. 
Mixing means an initially localized distribution can eventually spread to the whole phase space. 
Due to the linearity and absence of phase space in quantum mechanics,  quantum ergodicity and mixing are hard to define, and always discussed with ambiguous definition. 
Even though the classical counterparts of ergodicity and mixing can provide some hints, as discussed by Toda {\it et al.} in Ref.~\cite{Toda1989}, a quantum definition is still in need. Because quantum dynamics has it own features, and many systems, like spin systems, have no correspondence in classical mechanics. 

Our discussion in this section will center on the EE definitions of quantum ergodicity and mixing given in Ref.~\cite{zhang2016ergodicity}. 
These definitions are inherently quantum as they involve only eigen-energies: 
({\it i}) If there is no degeneracy in eigen-energies, the quantum system is ergodic. 
In this case, it can be shown that for a given observable its long time average is equal to its ensemble average.   
({\it ii}) If there is no degeneracy in the pairwise difference (or, loosely, gap) of
eigen-energies, the system is mixing as one can show that the fluctuation of a given observable is small.
However, it is still not clear whether these two definitions are consistent with our intuitive understanding of ergodicity and mixing in terms of dynamics in phase space. 
Here we use QKR to illustrate their consistency,  and show some special feature in dynamics of QKR.

For QKR, its Hamiltonian changes periodically with time, so there is no energy eigenstate. 
However, as is well known, for a periodically driven system, Floquet states play the role of eigenstates,  
while quasi-energies play the role of eigen-energies~\cite{Shirley,Luo}.
Therefore, a periodically driven quantum system is ergodic if there is no degeneracy in the quasi-energies and it is mixing when there is no degeneracy in the gaps of quasi-energies. 
One can similarly prove that the former leads to the long time average of an observable being equal to its ensemble average 
and the latter implies small fluctuation of an observable. The detail of the proof can be found in Appendix~\ref{proof}. 

We use the two parameters $\eta$ and $\zeta$ introduced in Ref.~\cite{zhang2016ergodicity} to 
measure the degeneracy of quasi-energies and their gaps, respectively. 
To numerically compute $\eta$, we evenly divide the whole quasi-energy range into $M$ small intervals and compute $\eta$ as
\be
\eta=N\sum_{i=1}^{M}\left(\frac{b_i}{ N}\right)^2\,,
\ee
where $b_i$ is the number of quasi-energies falling in the $i$th interval. 
Because the quasi-energies fall in $ [0, 2\pi) $ and have period $ 2\pi $, we define the difference between quasi-energies $ E_i, E_j $ as 
\be
\Delta E_{ij} = \min\{|E_i-E_j|, 2\pi - |E_i-E_j| \},\, \Delta E_{ij} \in [0, \pi)\,.
\ee
Similarly, we have
\be
\zeta=\frac{N(N-1)}{2}\sum_{i=1}^{M}\left(\frac{c_i}{ N(N-1)/2}\right)^2\,,
\ee
where $c_i$ is the number of quasi-energy gaps falling in the $i$th interval.  
Note that $ \eta, \zeta$ increase with the degeneracy of quasi-energies (gaps), and have minimum $ 1 $ when none of these quasi-energies (gaps) falls into the same interval.
In numerical computation, $M$ should be within a proper range. 
In the Appendix \ref{diff}, we explain in detail how these two parameters $\eta$ and $\zeta$ are computed  
and interpreted in a different way from Ref.~\cite{zhang2016ergodicity}. 

Analogous to the period of momentum in CKR, $\hat{U}_{m,n}$ is invariant under the transformation $m, n\rightarrow m+N, n+N$ when $N$ is even. Therefore, to facilitate the computation of Floquet states and quasi-energies and avoid numerical problems, we can apply a periodical condition in $ l $ space with period $ N $.
Our numerical results for the two degeneracy parameters $\eta$ and $\zeta$ at different  $\he = 2\pi/N$
are shown in Fig.~\ref{comp}~(a)(b). They have very similar behavior: the degeneracy decreases as 
the kicking strength  increases. 
Specifically, for small $ K $, although the curves shift slightly 
for different $N$, both $\eta$ and $\zeta$ drop quickly and reach a plateau.
As the kicking strength $K$ further increases, curves for different $N$ converge on each other. 
The second sharp drop happens around $ K = 1 $, which coincides with the transition point to chaos in the corresponding classical dynamics. The KAM theorem shows that the classical system becomes chaotic at $ K_c = 0.971635$~\cite{mackay1985converse,ChirikovScholarpedia}. 
For $ K $ larger than $1$, $\eta$ and $\zeta$ have reached their minimum $1$, which means there is no degeneracy of eigen-energies (gaps).

The transition of quantum dynamics indicated by the behaviors of $\eta$ and $\zeta$ in Fig.~\ref{comp} 
is supported by more intuitive numerical results. We choose an initial state that is localized 
in a single Planck cell, and then compute how it evolves with time. The numerical results are
shown in Fig.~\ref{comp}~(c)(d) for two different kicking strengths $K=0.5$ and $K=5$. 
At $K=0.5$, where both $\eta$ and $\zeta$ are large and the degeneracy is high, 
the wave packet does not spread much in quantum phase space. 
In contrast, at $K=5$, where both $\eta$ and $\zeta$ are small and the degeneracy is low, the wave packet spreads over almost all phase space after only $14$ kicks, which is a clear and intuitive indication of ergodicity and mixing. 
Apart from these snapshots, these different dynamical behaviors can be more comprehensively captured with the quantum entropy defined in Ref.~\cite{han2015entropy}. For QKR, this quantum entropy is
defined as
\be
 \mathcal{S}(t)=-\sum_{\px,\pp}P_{\px,\pp}\ln P_{\px,\pp}\,,
\ee
where $P_{\px,\pp}$ is the probability  of the state being at $\ket{\px,\pp}$. The dynamical evolution of this 
entropy for different kicking strength is shown in Fig.~\ref{comp}~(e). 
Periodical condition in momentum is also applied here to reduce the computational burden at large $ K $, so the entropy will saturate.
For kicking strength $K = 0.5$, $ \mathcal{S} $ remains small for all starting states, which means that the quantum states stay localized. 
For $K = 5$, $ \mathcal{S} $ grows quickly, representing the quantum states spread to the whole phase space. For kicking strength $K_c=0.972$, the behavior of $\mathcal{S}$ greatly depends on the initial states, so its mean value are close to $K=0.5$ but has a much larger variation.

In the last section we have shown that quantum dynamics of a kicked rotor in phase space can be reduced
to its classical counterpart when $N$ is very large or $\he$ is very small. 
This correspondence allows one to define a QKR as ergodic (or mixing) when its classical counterpart is ergodic (or mixing). This was in fact tried in Ref.~\cite{Toda1989} with the Q representation. 
The advantage of  their approach is that so-defined quantum ergodicity and mixing naturally reflect
how we understand ergodicity and mixing in classical dynamics. The disadvantage is that the definitions
are not inherently quantum mechanical and they are hard to be applied to general quantum systems, for 
example, spin systems. In contrast, the EES definitions of quantum ergodicity and mixing 
involve only eigen-energies and, therefore, inherently quantum mechanical.
With the results in Fig.~\ref{comp}, along with Fig.~\ref{phase}, we have shown with kicked rotor that these 
two definitions agree with each other in systems where they both can be applied. 

In Fig.~\ref{comp}~(a)(b),  we notice that  the size of $ \he $ (equivalently, $N$) has no impact with large $ K $, 
but have a nontrivial effect with small kicking strength. This may imply that more and more non-local 
conserved quantities are revealed as 
the system size $N$ increases, leading to a decrease  in quantum ergodicity parameter $\eta$ and 
mixing parameter $\zeta$.  The reason for the plateau
at small $K$ in these parameters is still not clear. We guess that this plateau is related to the generation of partial chaos.

\begin{figure}
\centering
\hspace*{-0.4cm}
\includegraphics[width=1.05\linewidth]{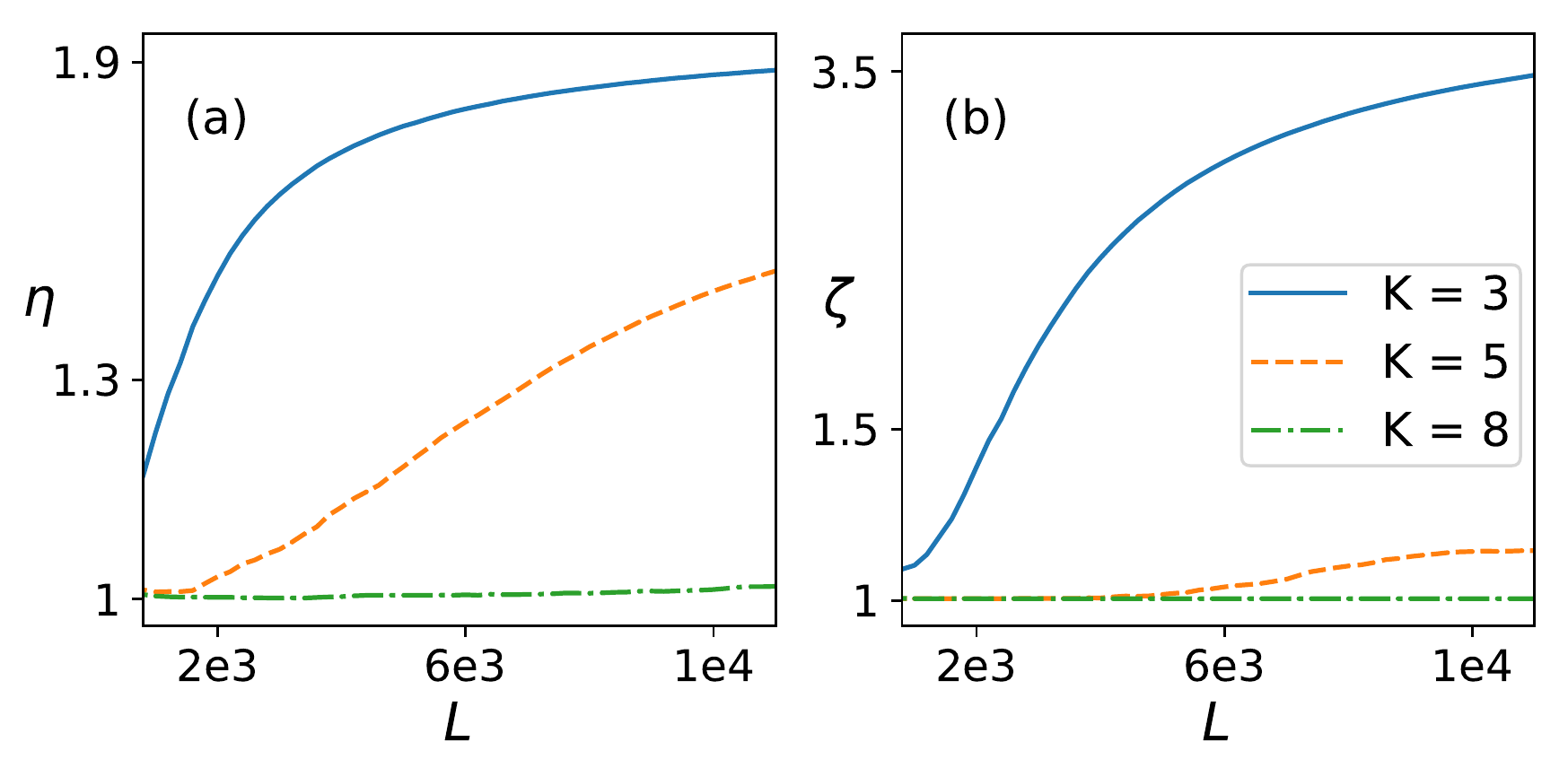}
\caption{\label{local} (color online) $ \eta, \zeta $ in EE definitions as a function of the number $ L $ of Floquet states. $ \he = 4 \pi / 53 \sqrt{5} $. y axis is in the log scale. x axis starts at $ 1e3 $ to get enough quasi-energies for statistics.}
\end{figure}

We have emphasized that our study focuses on short time scales, $t<t_H$.  In QKR, 
the rotor will explore higher momentum states with more kicks  in diffusion region.  Therefore, 
time scale  and the size of the Hilbert space  are related. 
To be specific, Heisenberg time $t_H$ relates to localization length in Hilbert space. That is to say, when one observe dynamical localization in time evolution, it also implies large degeneracy in quasi-energy levels of the Floquet operator with corresponding system size. Hence, if our definition of $\eta$ and $\zeta$ are good enough, they can capture the signal of localization. In our previous discussion, the system size is small, which ensures we are studying short time limit and it is appropriate to set periodical boundary condition.  
Now we discuss briefly longer time scale $t \gtrsim t_H$ by extending the allowed range of Floquet states  without periodical boundary condition. With a given center state $ \ket{k_0} $, we compute the $ L $ Floquet states that is closest to the center state, 
and the corresponding parameter $ \eta, \zeta $.  Since we are only interested in chaotic region, 
so we focus on cases $ K > K_c $.  For small system size $ L $, $ \eta$ and $\zeta $ are close to $ 1 $, suggesting the short time dynamics is ergodic and mixing. Quantum effects starts to appear as increasing $\eta$ and $\zeta$ with  increasing $ L $, as shown in Fig.~\ref{local}. This phenomenon is due to increasing level degeneracy because of localization effect for large system size.
Accordingly, the level statistics changes from Wigner-Dyson distribution to Poisson distribution.
 As the localization length of the system increases with $ K $, as predicted in Ref.~\cite{Izrailev90},  the deviation of $\eta$ and $\zeta$ from 1 are postponed for increasing K.
Dynamical localization is a pure quantum effect, so the results with  EE definitions in Fig.~\ref{local}
demonstrate the power of our method.

\section{Conclusion}
In sum, we have demonstrated quantum-classical correspondence of ergodicity and mixing with QKR. 
Such a correspondence was established with two very different approaches. 
The first approach used only the Floquet quasi-energies of QKR and be easily generalized to 
any quantum system, including quantum spin system. In the second approach, 
with a method originated from von Neumann, we were able to project 
wave functions unitarily to quantum phase space and plot quantum Poincar\'e section. 
It allowed us to examine quantum ergodicity and mixing in a way very similar to the classical approach. 
The results obtained with both approaches are consistent with each other, and also with the classical results.

\acknowledgements
This work was supported by the The National Key Research and Development Program of China (Grants No.~2017YFA0303302) and the National Natural Science Foundation of China (Grants No.~11334001, No.~11429402 and No.~11734010).

\appendix

\section{Localization of SFME basis} \label{app_wave_packet}
\def\noo{\nonumber}
In the main text, a set of orthonormal and complete basis is introduced and used to project
wave function unitarily onto quantum phase space. These basis wave functions $\ket{\px,\pp}$
are defined as a superposition of finite momentum eigenstates (SFME).  Here we examine
how localized these SFME wave packets are.  For this purpose, we only need to 
examine one wave packet $\ket{0,0}$ as other FSME wave packets can be obtained
by translation (see Eq.~(\ref{xp})).  

The wave packet $\ket{0,0}$ is plotted in  Fig.~\ref{packet}~(a), where the localization is quite obvious. 
To get more quantitative understanding, we compute the spread of $ \ket{0, 0} $ as  
\begin{align}
\var(\ll) &= \avera{(\ll - \avera{\ll})^2} = \frac{1}{\ell}\sum_{i = 1}^\ell (\frac{\ell+1}{2}-i)^2 = \frac{\ell^2-1}{12} \\
\var(\tt) &= \avera{(\tt - \avera{\tt})^2} = \int_{0}^{2\pi}\tt^2\left|\braket{\tt|0, 0}\right|^2d\tt \noo \\
&= \frac{\pi^2}{3}+\frac{4}{\ell}\sum_{k=1}^{\ell-1}\frac{(-1)^k(\ell-k)}{k^2}
\end{align}
It is clear that the spread $ \var(\tt) $ converges to $ 0 $ asymptotically as $ 1 / \ell $ while $ \var(\ll) $ diverges as
$\ell^2$.  However, what is important is the relative spreads $\sqrt{\var(\tt)}/N$ and $\sqrt{\var(\ll)}/N$. 
As in the main text, we choose $N = \ell ^ 2$. Then  the relative 
spread on $ \tt $ and $ k $ converge as $ \ell ^ {-1} $ and $ \ell ^ {-2.5} $, respectively, as shown in Fig.~\ref{packet}~(b). 

\makeatletter 
\renewcommand{\thefigure}{A\@arabic\c@figure}
\makeatother
\setcounter{figure}{0}

\begin{figure}
\hspace*{-0.5cm}
\centering
\includegraphics[width=1.06\linewidth]{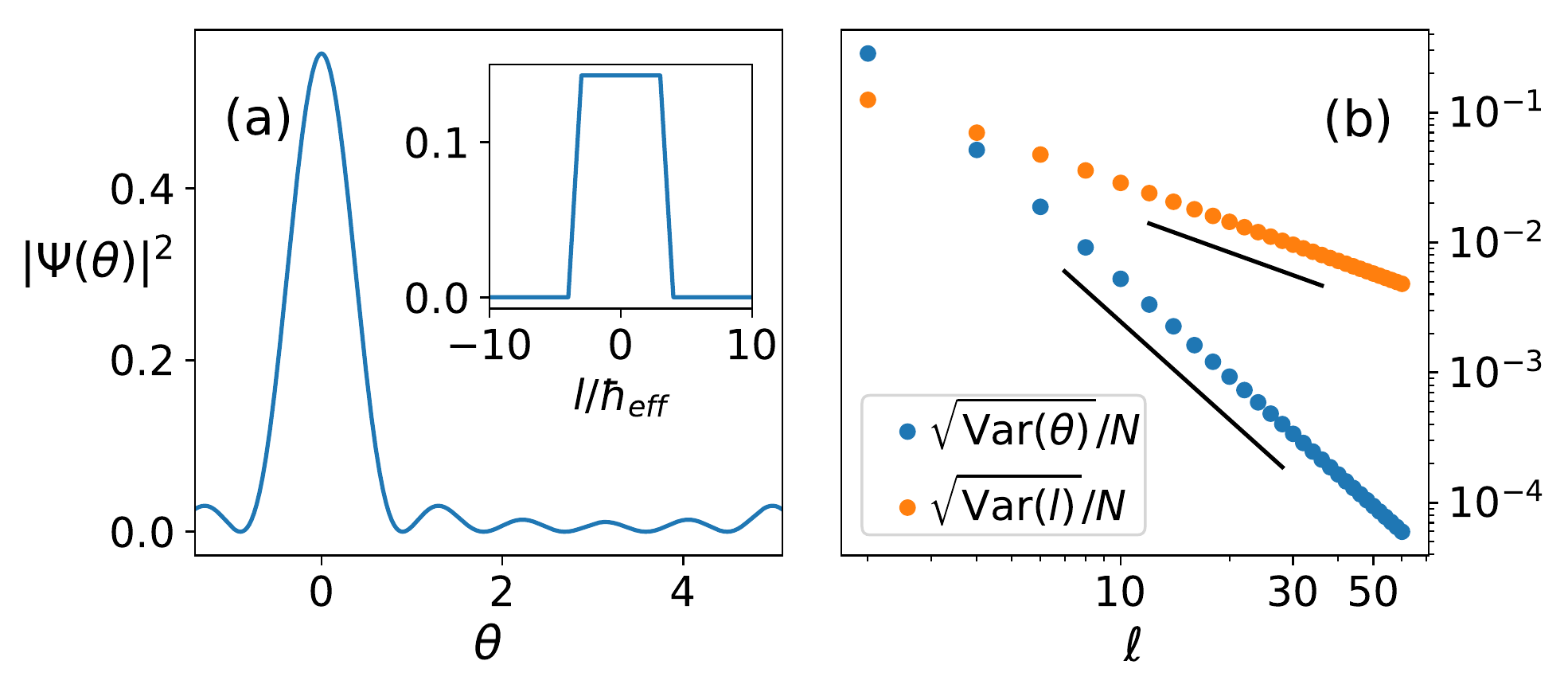}
\caption{\label{packet} (a) The distribution of $ \ket{0, 0} $ with $ \ell = 7 $ on angle  and angular momentum (inset). (b) The relative spread of $ \tt $ and $ k $ with $ \ell $. Two reference line have slope $ -1 $ and $ -2.5 $. }
\end{figure}

\section{Proof of ergodicity and mixing in periodical driven system}\label{proof}

\paragraph{Ergodicity}

\def\pst{\Psi (t)}
\def\fln{\phi_n (t)}
\def\flm{\phi_m (t)}
For a periodical Hamiltonian $ \mathcal{H} $ with Flouqet states $ \ket{\phi_n(t)} $, any state can be represented as 
\be
\ket{\Psi(t)} = \sum_n c_n\ket{\phi_n(t)}
\ee

\def\ah{\hat{A}}
\def\amn{A_{mn}(t)}
\def\anm{A_{nm}(t)}
\def\que{E}
\newcommand{\avt}[1]{\braket{#1}_T}
\newcommand{\ave}[1]{\braket{#1}_E}
\newcommand{\avtau}[1]{\left< #1 \right>_{\tau}}
For an observable $ \ah $, its expectation over $ \ket{\pst} $ is 
\be
\braket{\pst|\ah|\pst} = \sum_{m, n}c_m^* c_n A_{mn}(t)\,,
\ee
where $ A_{mn}(t) = \braket{\flm|\ah|\fln} $. By definition of Floquet states, the long time average for $ \amn $ is 
\be
\avt{\amn} = \frac{\avtau{\amn}}{[T / \tau]} \sum_{a = 0}^{[T / \tau]-1} e^{-i (\que_n - \que_m)a/\hbar}\,.
\ee
The contribution after $ \tau [T / \tau] $ is omitted because it would vanish for large $ T $. When the ergodic condition is satisfied, using the fact that when $ k \in [0, 2\pi) $,
\be
\lim_{N \to \infty}\frac{1}{N}\sum_{n = 1}^N e^{ink} = 
\begin{cases}
1 & k = 0\\
0 & k \neq 0 
\end{cases}\,,
\ee
therefore $ \avt{\amn} = 0 $ for $ m \neq n $. That is, 
\be
\avt{\ah} = \sum_m c_m^*c_m \avtau{A_{mm}(t)} = \ave{\ah}\,.
\ee
Only diagonal terms persists, which means that the long time average equals to the ensemble average $ \ave{\ah} $. 

\def\alk{A_{lk}(t)}

\paragraph{Mixing}

To quantify the fluctuation, we compute the variance of $ \ah $
\begin{align}
\avt{\sigma^2_A} &= \avt{\braket{\ah}^2} - \ave{A}^2 \noo \\
&= \sum_{k, l, m, n}\rho_{kl}^*\rho_{mn} \avt{\amn\alk} \noo \\
&- \sum_{m, n} \rho_{mm}\rho_{nn} \avtau{A_{mm}(t)}\avtau{A_{nn}(t)}\,,
\end{align}
where $ \rho_{mn} = c_m^*c_n $. 
If we omit the contribution after $ \tau [T / \tau] $, we have
\begin{align}
&\avt{\amn\alk}\noo \\
=& \frac{\avtau{\amn\alk}}{[T / \tau]} \sum_{a = 0}^{[T / \tau]-1} e^{-i [(\que_n - \que_m) - (\que_l - \que_k)] a/\hbar}\,.
\end{align} 
If the condition of mixing is satisfied, similarly we have
\be
\avt{\amn\alk} = 
\begin{cases}
\avtau{\amn\alk} & \delta_{mn}\delta_{kl} = 1, \\
&  \delta_{mk}\delta_{nl} = 1\\
0 & \textrm{otherwise}
\end{cases}
\ee
Therefore, the fluctuation of $ \ah $ is bounded by 
\begin{align}
\avt{\sigma^2_A} &= \sum_{m\neq n} \rho_{mn}^* \rho_{mn}\avtau{\amn\anm} \noo \\
&\leq \sum_{m, n} \rho_{mm} \rho_{nn}\avtau{\amn\anm} \noo \\
&= \avtau{\Tr (\rho_{mc} A A^\dagger \rho_{mc})}\,,
%&\leq \sqrt{\avtau{}}
\end{align}
where $ \rho_{mc} $ is the diagonal matrix with $n$th element being $ \rho_{nn} $. By the Cauchy-Schwartz inequality, we have
\begin{align}
\Tr (\rho_{mc} A A^\dagger \rho_{mc}) &\leq \sqrt{\Tr (AA^\dagger \rho_{mc}^2) \Tr (A^\dagger A\rho_{mc}^2) } \noo \\
&\leq \norm{AA^\dagger} \Tr \rho_{mc}^2 \,,
\end{align}
where $ \norm{AA^\dagger} = \sup \{ \braket{\braket{\Psi|AA^\dagger|\Psi}}_\tau: \ket{\Psi} \in \mathscr{H} \} $ is the upper bound for the average of $ AA^\dagger $ in the Hilbert space $ \mathscr{H} $. Finally, we have for the fluctuation
\be
F_A^2 \equiv \frac{\avt{\sigma^2_A}}{\norm{AA^\dagger}}\leq \Tr \rho^2_{mc}
\ee

\section{Quantification of the degeneracies in eigen-energies} \label{diff}
We have presented numerical results in the main text for two parameters, $\eta$ and $\zeta$, 
which characterize the degeneracies of Floquet quasi-energies. As $\eta$ and $\zeta$ mathematically are the same,
we focus only $\eta$ and show in detail how it is computed. 

\begin{figure}
\includegraphics[width=0.8 \linewidth]{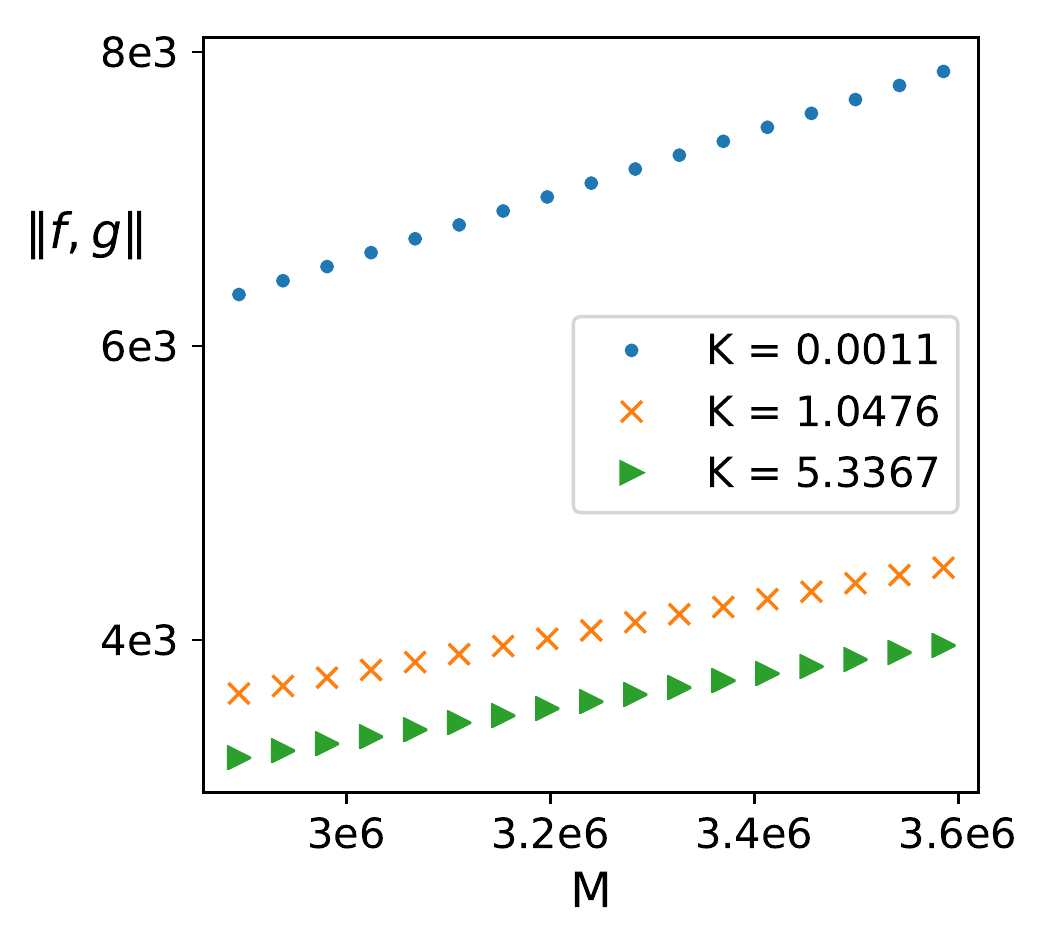}
\caption{\label{linear_dist} Growth of $ \norm{f, g} $ with $ M $ when $ \ell = 120 $\,.}
\end{figure}

Our task is to define a measure for the degeneracy of $ N $ quasi-energies distributed 
in the interval $ [ 0, 2\pi] $. The major difficulty is that these quasi-energies are obtained  
numerically, and therefore there is no rigorous degeneracy. The degeneracy here only means that  
some quasi-energies lie very close to each other. The more clustered the distribution is,  
the more it deviates from a uniform distribution on this interval.  To quantify this deviation,
we define the distance between the empirical distribution $f(x)$ and the uniform distribution $g(x)$ as 
\be
\norm{f(x),g(x)} \equiv \int_0^{2\pi} \big[f(x)-g(x)\big]^2dx\,.
\ee
To compute it, we divide this interval equally to $ M $ parts and the  distance becomes 
\begin{align}
d(M) =\norm{f(x),g(x)} &= \sum_{i=1}^{M}(\frac{b_i M}{2\pi N}-\frac{1}{2\pi})^2\frac{2\pi}{M}\\
&= \frac{M}{2\pi}\sum_{i=1}^{M}\left(\frac{b_i}{N}\right) ^ 2 -\frac{1}{2\pi}\,,
\end{align}
where $ b_i $ is the number of quasi-energies in the $i$th interval. Apparently, 
the distance $d(M)$ is a function of $M$. 

Degeneracy means some gaps between 
quasi-energies are significantly smaller than others. As a result, we expect that within a proper range 
of divide number $ M $ only degenerate quasi-energies stay in the same interval. 
This means that $d(M)$ is a linear function of $M$ within this proper range and $\eta$ emerges 
as the slope of this function (up to a constant multiple $N$).  
Examples of this linear relation are shown in Fig.~\ref{linear_dist}. In our 
numerical computation, we usually choose $ M \sim 200 N $.  The parameter $\eta$ is extracted as the slope
normalized with respect to  $1/N$  to make the minimal of $\eta $ to be $ 1 $.
%\bibliography{mixing}

\end{document}